\documentclass[twocolumn,showpacs,preprintnumbers]{revtex4}
\usepackage{amsmath}
\usepackage{graphicx}

\begin{document}

\title{Superfast convergence effect in large orders of the perturbative and 
$\varepsilon$ expansions for the $O(N)$-symmetric $\phi^{4}$ model.}
\author{P.V. Pobylitsa}
\affiliation{Petersburg Nuclear Physics Institute, Gatchina, St.~Petersburg, 188300,
Russia}

\begin{abstract}
Usually the asymptotic behavior for large orders of the perturbation theory
is reached rather slowly. However, in the case of the $N$-component 
$\phi^{4} $ model in $D=4$ dimensions one can find a special quantity that
exhibits an extremely fast convergence to the asymptotic form. A comparison
of the available $5$-loop result for this quantity with the asymptotic value
shows  agreement at the $10^{-3}$ level. An analogous superfast
convergence to the asymptotic form happens in the case of the $O(N)$-symmetric
anharmonic oscillator where this convergence has inverse factorial type.
The large orders of the $\varepsilon$ expansion for
critical exponents manifest a similar effect.
\end{abstract}

\maketitle

\section{Introduction}

\label{Introduction-section}

The series of the perturbation theory $\sum\limits_{k}a_{k}g^{k}$ usually
have factorially growing coefficients \cite{BW-69,BW-73,Lipatov-77b,BLZ-77}: 
\begin{equation}
a_{k}=ck!A^{k}k^{b}\left[ 1+O(k^{-1})\right] \,.  \label{a-k-asymptotic}
\end{equation}
One of the problems with these series (see reviews
\cite{KP-2002,Suslov-05,CMRSJ-07} and references therein) is the slow convergence
of the coefficients $a_{k}$ to asymptotic form (\ref{a-k-asymptotic}). In
particular, for the $O(N)$-symmetric $\phi ^{4}$ model,
\begin{equation}
L=\frac{1}{2}\sum\limits_{a=1}^{N}\left( \partial _{\mu }\phi _{a}\right)
^{2}+\frac{1}{4!}g\left( \sum\limits_{a=1}^{N}\phi _{a}^{2}\right) ^{2}\,,
\end{equation}
the agreement between available 5-loop results \cite{KNSCL-91} and
asymptotic formulas (\ref{a-k-asymptotic}) is rather poor, which may be
attributed to the large $O(k^{-1})$ correction \cite{Kubyshin-84}.

The aim of this paper is to attract attention to some \emph{special
quantities} whose asymptotic behavior is reached much faster
than  the typical slow $O(k^{-1})$ convergence in eq. (\ref{a-k-asymptotic}).
In particular, it will be shown that the 5-loop $\beta $
function of the $N$-component $\phi ^{4}$ model contains some properties
which agree with the asymptotic behavior with relative accuracy better
than $10^{-3}$.

In the case of the perturbative $\beta $ function of the $O(N)$-symmetric
$\phi ^{4}$ model $\beta (g,N)=\sum_{k=2}^{\infty }\beta _{k}(N)g^{k}$, the
large-$k$ asymptotic behavior of $\beta _{k}(N)$ is given by
\begin{equation}
\beta _{k}(N)\overset{k\rightarrow \infty }{\longrightarrow }
\beta _{k}^{\mathrm{as}}(N)=c_{N}\mathrm{\,}k!\left( -16\pi ^{2}\right) ^{-k}
k^{\frac{N}{2}+3}\,  \label{beta-asymptotic}
\end{equation}
with $c_{N}$ computed in Ref. \cite{BLZ-77} and simplified in 
Refs. \cite{Kubyshin-84,MW-78,MWA-1984}:
\begin{gather}
c_{N}=\left( \frac{9}{2\pi }\right) ^{\frac{N-1}{6}}\,\exp \left\{ (N+2)
\left[ \frac{\zeta ^{\prime }(2)}{\pi ^{2}}-\frac{\gamma }{6}-\frac{1}{4}
\right] -3\right\}  \notag \\
\times 96^{3/2}\frac{(N+8)\Gamma \left( \frac{5}{2}\right) }{9\Gamma \left( 
\frac{N}{2}+2\right) }\int_{0}^{\infty }dx\,x^{(N+17)/3}\left[ K_{1}(x)
 \right] ^{4}.  \label{c-N-phi-4}
\end{gather}

Instead of the direct comparison of $\beta_{k}^{\mathrm{as}}(N)$ with 
$\beta_{k}(N)$ (which shows rather poor agreement \cite{DKT-78}) we choose
another way. At $k\geq3$, $\beta_{k}(N)$ is a polynomial in $N$ of degree 
$k-2$ and has $k-2$ roots
\begin{equation}
\beta_{k}(N_{k,r})=0\quad(1\leq r\leq k-2)\,.
\end{equation}
Using the 5-loop $\beta$ function of the $N$-component $\phi^{4}$
model in the MS scheme \cite{KNSCL-91}, one can easily compute
these roots numerically. The values are listed in Table \ref{table-1}.

\begin{table}[ptb]
\begin{tabular}{||l|l||}
\hline\hline
$k$ & $N_{k,r}$ \\ \hline\hline
$3$ & $-4.66667$ \\ \hline
$4$ & $-4.02495,\,-41.3989$ \\ \hline
$5$ & $-4.01968,\,-12.0757,\,3218.75$ \\ \hline
$6$ & $-4.00173,\,-8.75514,\,-44.0331,\,504.74$ \\ \hline\hline
\end{tabular}
\caption{Roots $N_{k,r}$ of polynomials $\protect\beta_{k}(N)$.}
\label{table-1}
\end{table}

In this set of roots one can clearly see a magic sequence
\begin{equation}
\left\{ -4.6667,\,-4.0249,\,-4.0197,\,-4.0017,\ldots \right\} \rightarrow
-4\,  \label{root-sequence-phi-4}
\end{equation}
which seems to converge very fast to the value $-4$. The aim of this paper
is to show that we deal with a rather interesting effect of \emph{superfast
convergence} which is closely related to the \emph{factorial divergence} of
large-order perturbative coefficients (\ref{a-k-asymptotic}),
(\ref{beta-asymptotic}).

In Sec.\ \ref{oscillator-section} we will study a similar phenomenon in a
much simpler model of the $O(N)$-symmetric anharmonic oscillator. This model
will allow us to understand the origin of the superfast convergence and its
relation to the properties of asymptotic expressions at the nonphysical
point $N=-4$. In Sec.\ \ref{convergence-origin-section} we return to the 
$\phi^{4}$ model and show that the situation is quite similar. 
In Sec.\ \ref{epsilon-expansion-section} an analogous effect of superfast convergence is
checked in large orders of the $\varepsilon$ expansion for critical
exponents.

\section{Anharmonic oscillator}

\label{oscillator-section}

The $N$-component anharmonic oscillator
\begin{equation}
H_{N}=\frac{1}{2}\sum\limits_{a=1}^{N}\left( p_{a}^{2}+x_{a}^{2}\right)
+g\left( \sum\limits_{a=1}^{N}x_{a}^{2}\right) ^{2}\,
\label{H-quartic-oscillator}
\end{equation}
and the properties of the perturbative expansion of the energy of the ground
state
\begin{equation}
E(g,N)=\sum\limits_{k=0}^{\infty }E_{k}(N)g^{k}\,  \label{E-expansion}
\end{equation}
are well studied \cite{BW-69,BW-73,SZ-79,Zinn-Justin-81,ZJ-04,CMRSJ-07}. 
$E_{k}(N)$ is a polynomial of degree $k+1$ with $k+1$ roots $\nu _{k,r}$:
\begin{equation}
E_{k}(\nu _{k,r})=0\,.  \label{P-nu-root-eq}
\end{equation}
The roots close to $-4$ are listed in Table \ref{table-2}, we label them
with $r=1$. They were computed numerically \cite{PP-08} using methods of
Refs. \cite{BW-69,BW-73,SZ-79,Zinn-Justin-81}. At $k\geq 10$, the roots $\nu
_{k,1}$ are real and exhibit a very fast {\em inverse factorial convergence} to $-4$ which
is described by the asymptotic formula derived in Appendix (see also
Ref.~\cite{PP-08}):
\begin{equation}
\nu _{k,1}+4\overset{k\rightarrow \infty }{\longrightarrow }12
\sqrt{2\pi }\frac{(-1)^{k}}{k!}3^{k/2}k^{3/2}\,.  \label{nu-k-asymptotic}
\end{equation}

\begin{table}[ptb]
\begin{tabular}{||l|l||l|l||}
\hline\hline
$k$ & $\nu_{k,1} (\nu_{k,1}^{\ast})$ & $k$ & $\nu_{k,1}$ \\ \hline\hline
$5$ & $-3.22834\pm i0.426293$ & $12$ & $-4+0.0017843$ \\ \hline
$6$ & $-3.44545$ & $13$ & $-4-0.00027422$ \\ \hline
$7$ & $-3.63083\pm i0.34226$ & $14$ & $-4+0.00003787$ \\ \hline
$8$ & $-3.76443$ & $15$ & $-4-4.8625\times10^{-6}$ \\ \hline
$9$ & $-3.9583\pm i0.226557$ & $\ldots$ &  \\ \hline
$10$ & $-4+0.0423159$ & $30$ & $-4+2.61370\times10^{-22}$ \\ \hline
$11$ & $-4-0.0123126$ & $31$ & $-4-1.53497\times10^{-23}$ \\ \hline\hline
\end{tabular}
\caption{Roots $\protect\nu_{k,1}$ of polynomials $E_{k}(N)$ approaching the
value $-4$.}
\label{table-2}
\end{table}

\begin{figure}[ptb]
\begin{center}
\includegraphics[
height=3.2785in,
width=3.3944in
]{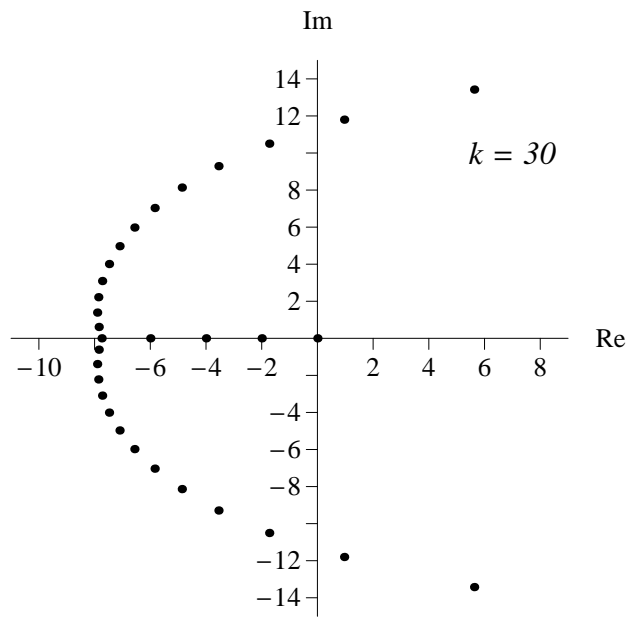}
\end{center}
\caption{Distribution of roots $\protect\nu_{30,r}$ of polynomial 
$E_{30}(N)$ in the complex plane. The roots approaching the negative even
points $-4,-6,-8$ are clearly seen as well as the stable roots $0$ and $-2$.}
\label{orig-fig-1}
\end{figure}

The explanation of this sequence of roots tending to $-4$ is simple. The
large-$k$ asymptotic formula for $E_{k}(N)$ with the $O(k^{-1})$ correction 
\cite{Zinn-Justin-81} is
\begin{align}
& E_{k}(N)\overset{k\rightarrow \infty }{=}(-1)^{k+1}\Gamma 
\left( k+\frac{N}{2}\right) 3^{k+\frac{N}{2}}
\frac{2^{N/2}}{\pi \Gamma \left( N/2\right) } 
\notag \\
& \times \left[ 1-\frac{1}{6k}\left( \frac{5}{3}+\frac{9}{2}N+\frac{7}{4}
N^{2}\right) +O(k^{-2})\right] \,.  \label{E-k-N-asymptotic}
\end{align}
This asymptotic formula contains a factor of $\left[ \Gamma (N/2)\right]
^{-1}$ that has zeros at $N=0,-2,-4\ldots $ At $k\geq 1$, all polynomials 
$E_{k}(N)$ have a factor of $N(N+2)$ so that at we have common zeros of 
$\left[ \Gamma (N/2)\right] ^{-1}$ and $E_{k}(N)$ at $N=0,-2$. The zero of 
$\left[ \Gamma (N/2)\right] ^{-1}$ at $N=-4$ corresponds to the \emph{fast
convergent sequence of zeros} of $E_{k}(N)$: $\nu _{r,1}\rightarrow -4$. One
can wonder what happens with large zeros $N=-6,-8,\ldots $ of the function 
$\left[ \Gamma \left( N/2\right) \right] ^{-1}$ appearing on the RHS of
(\ref{E-k-N-asymptotic}). It can be checked numerically that at larger values of 
$k$ polynomials $E_{k}(N)$ get additional series of roots convergent to
the points $N=-6,-8,\ldots $. As an example, the zeros of the polynomial 
$E_{30}(N)$ are shown in Fig. \ref{orig-fig-1} where the roots approaching
points $N=-6,-8$ are clearly seen.

\section{Origin of the roots convergent to $-4$ in the $\protect\phi^{4}$
model}

\label{convergence-origin-section}

Now we return to the field theory. Expression (\ref{c-N-phi-4}) contributes
a factor of $\left[ \Gamma \left( 2+\frac{N}{2}\right) \right] ^{-1}$ to 
$\beta _{k}^{\mathrm{as}}(N)$. Therefore $\beta _{k}^{\mathrm{as}}(N)$
vanishes at $N=-4,-6,\ldots$ (starting from $N=-8$ we must take into account the
divergence of the integral containing $\left[ K_{1}(x)\right] ^{4}$ and the
extra factor of $(N+8)$). These zeros of $\beta _{k}^{\mathrm{as}}(N)$ play
the same role as zeros of $\left[ \Gamma \left( N/2\right) \right] ^{-1}$ in
eq. (\ref{E-k-N-asymptotic}) for the anharmonic oscillator. The zeros close
to $N=-4$ appear already in the lowest orders of the $\beta $ function as we
saw in (\ref{root-sequence-phi-4}).

Strictly speaking, asymptotic behavior (\ref{beta-asymptotic}),
(\ref{c-N-phi-4}) was derived for the function $\beta $ in the MOM scheme
whereas the 5-loop $\beta$ function was computed in Ref. \cite{KNSCL-91}
in the MS scheme. The question about the asymptotic behavior of the
perturbation theory for the $\beta$ function in the MS scheme was considered
in Ref. \cite{KP-2002}. 
The MOM $\beta$ function is available only in
four loops \cite{DKT-78}.  For the MOM $\beta$ function, we find
a similar sequence of roots for 2, 3 and 4 loops
$\{-4.66667,\,-4.09365\,,\,-4.12669\}$. We see a proximity to $-4$ but in the
4-loop MOM case the convergence is seen less clearly than in the 5-loop MS
sequence (\ref{root-sequence-phi-4}).

One can trace the origin of the factor
$\left[ \Gamma \left( 2+(N/2)\right)\right] ^{-1}$ in the derivation of
asymptotic expressions
(\ref{beta-asymptotic}), (\ref{c-N-phi-4}) in Ref.~\cite{BLZ-77}. In a more
general context of the $O(N)$-symmetric $\left( \phi ^{2}\right) ^{M}$ model
in $D=2M/(M-1)$ dimensions this factor has the form $\left[ \Gamma \left(
M+(N/2)\right) \right] ^{-1}$ and comes from the integration over the $O(N)$
collective coordinates.

\section{$\protect\varepsilon$ expansion}

\label{epsilon-expansion-section}

A similar effect of the fast convergence of roots to the value $-4$ can be
found in large orders of the $\varepsilon $ expansion for critical
exponents. Using a compact notation for the critical exponents $z=\omega
,\nu ^{-1},\eta $ of the universality class associated with the $O(N)$
symmetry, one can write the $\varepsilon $ expansion in the form 
\begin{equation}
z(\varepsilon ,N)=\sum_{k}\left[ \varepsilon (N+8)^{-2}\right]
^{k}P_{k}^{z}(N)\,.  \label{z-epsilon-expansion}
\end{equation}
Here $P_{k}^{z}(N)$ are polynomials of $N$. Using expressions for
$P_{k}^{z}(N)$ with $k\leq 5$ from Ref. \cite{KNSCL-91}, we can easily find
the roots
\begin{equation}
P_{k}^{z}(q_{z,k,r})=0
\end{equation}
tending to the value $-4$. These roots (labeled with $r=1$) are listed in
Table \ref{table-3}. The convergence to the value $-4$ is clearly seen. This
convergence is especially impressive for the critical exponent $\omega $.
Remember that this critical exponent is controlled by the $\beta $ function only: 
$\omega =2\beta _{\varepsilon }^{\prime }(g_{0})$, $\beta _{\varepsilon
}(g_{0})=0$. Therefore there must be a certain correlation with the extremely
good convergence that we saw for the roots associated with the coefficient
functions of the $\beta $ function in eq. (\ref{root-sequence-phi-4}).

\begin{table}[tbp]
\begin{tabular}{||l|l|l|l||}
\hline\hline
$k$ & $q_{\omega,k,1}$ & $q_{\nu^{-1},k,1}$ & $q_{\eta,k,1}$ \\ \hline\hline
$2$ & $-4.66667$ & $-3.38462$ &  \\ \hline
$3$ & $-3.96014$ & $-3.69220$ & $-4.49615$ \\ \hline
$4$ & $-4.02773$ & $-3.99515$ & $-3.79616$ \\ \hline
$5$ & $-3.99610$ & $-3.98743$ & $-4.04522$ \\ \hline\hline
\end{tabular}
\caption{Roots $q_{z,k,1}\rightarrow -4$ of polynomials $P_{k}^{z}(N)$ for
critical exponents $z=\protect\omega ,\protect\nu ^{-1},\protect\eta $.}
\label{table-3}
\end{table}

The appearance of the same limit value $-4$ in the case of the $\varepsilon $
expansion of critical exponents can be easily explained by the asymptotic
formulas derived in Ref. \cite{MWA-1984}:
\begin{equation}
P_{k}^{z}(N)=-\bar{C}_{z}k!\left[ -3\left( N+8\right) \right]
^{k}k^{(N+m_{z})/2}\left[ 1+O\left( \frac{\ln k}{k}\right) \right]
\end{equation}
where
\begin{align}
m_{\eta }& =6,\quad m_{\nu ^{-1}}=8,\quad m_{\omega }=10\,, \\
\bar{C}_{\nu ^{-1}}& =3\bar{C}_{\eta }\,,\quad \bar{C}_{\omega }
=\frac{3(N+8)}{N+2}\bar{C}_{\eta }\,,  \label{C-bar-omega}
\end{align}
\begin{align}
\bar{C}_{\eta }& =\frac{2^{-(N-16)/6}3^{(N+3)/2}\pi ^{-(N+8)/6}}{(N+8)\Gamma
\left( \frac{N}{2}+1\right) }\exp \left[ \frac{(N+2)\zeta ^{\prime }(2)}{\pi
^{2}}\right.  \notag \\
& \left. -\frac{1}{2}(N+6)\gamma -\frac{1}{4}(N+14)-\frac{3N+14}{N+8}\right]
\,.
\end{align}
Here we have the common factor $\left[ \Gamma \left( \frac{N}{2}+1\right) 
\right] ^{-1}$ that has zeros at $N=-2,-4,-6,\ldots $ The zero $N=-2$ is
trivial:

1) in the case of the critical exponent $\omega$, this zero cancels in expression
(\ref{C-bar-omega}) for 
$\bar{C}_{\omega}$,

2) in the case of critical exponents $\nu^{-1}$, $\eta$ the factor $(N+2)$ is
explicitly present in all orders of the $\varepsilon$ (apart from the
$\varepsilon^{0}$ contribution to $\eta$).

The zero at $N=-4$ is more interesting. It is responsible for the
convergent sequences $q_{z,k,1}\overset{k\rightarrow \infty }
{\longrightarrow }-4$.

\section{Conclusions}

In spite of the widespread skepticism about relevance of the large-order
asymptotic formulas for the current multi-loop calculations, we found
quantities that reach the asymptotic behavior with high precision in rather
low orders of the perturbation theory. The main idea was to turn from the
traditional analysis of the perturbation theory at fixed $N$ to the roots of
the $N$-dependence.

Several manifestations of the superfast convergence of these roots
to the asymptotic were studied:
anharmonic oscillator, perturbative beta function of the $\phi ^{4}$ model
and $\varepsilon $ expansion for critical exponents.
In the case of the anharmonic oscillator we could explicitly show that the
convergence of roots to the asymptotic value $-4$ is factorially fast.
In the $\phi^4$ model the situation is less clear: the argument
for the fast convergence of roots comes from the vanishing coefficient
in asymptotic formula (\ref{c-N-phi-4}) at $N=-4$.
It is rather astonishing that for the $\beta $ function and for the critical
exponent $\omega $ (directly related to the $\beta $ function) we have 
$10^{-3}$ agreement with the asymptotic value already in the 5-th loop. Note
that in the case of the anharmonic oscillator the convergence to the
asymptotic form starts later. This may be related to the fact that we have
complex roots in low orders of the perturbation theory for the anharmonic
oscillator whereas in the $\phi ^{4}$ field theory the roots are real from
the very beginning.

Although the superfast convergence deals with nonphysical values 
$N\rightarrow-4$, this effect may still have useful applications:

i) The proximity of the $N$-roots to $-4$ can be used as an error test in
multi-loop calculations.

ii) In field theory where the theoretical status of asymptotic formulas is
sometimes not reliable, the effect of the superfast convergence may play
the role of an additional control of theoretical assumptions.

iii) One could think about extensions of the method to other theories with a
polynomial dependence on various parameters.

\acknowledgments{I appreciate discussions with G.V.~Dunne, D.I.~Kazakov, N.~Kivel, 
L.N. Lipatov, V.Yu.~Petrov, I.M.~Suslov and A. Turbiner.}

\appendix                        

\section{Asymptotic estimates for the anharmonic oscillator}

\subsection{Large-$k$ asymptotic formula for the roots $\protect\nu_{k,r}$}

Let us derive asymptotic formula (\ref{nu-k-asymptotic}) for the roots of
equation $E_{k}(\nu_{k,1})=0$. In the limit $k\rightarrow\infty$, $\nu
_{k,1}\rightarrow-4$ we can solve this equation using a modified version of
asymptotic formula (\ref{E-k-N-asymptotic}) for the double limit 
$k\rightarrow\infty,N\rightarrow-4$:
\begin{equation}
E_{k}(N)\rightarrow(-1)^{k+1}\Gamma\left( k+\frac{N}{2}\right) 
3^{k+\frac{N}{2}}\frac{2^{N/2}}{\pi\Gamma\left( \frac{N}{2}\right) }+E_{k}(-4)\,.
\label{E-k-asymptotic-modified}
\end{equation}
We dropped the $1/k$ correction present in eq. (\ref{E-k-N-asymptotic})
but added the extra term $E_{k}(-4)$ which becomes essential at 
$N\rightarrow -4$. At $N\rightarrow-4$ we can simplify
eq. (\ref{E-k-asymptotic-modified})
\begin{equation}
E_{k}(N)\rightarrow (-1)^{k+1}\frac{k!}{k^{3}}\frac{3^{k-2}}{4\pi }\left(
N+4\right) +E_{k}(-4)\,.
\end{equation}
Now we find the roots of $E_{k}(N)$: 
\begin{equation}
\nu _{k,1}+4\overset{k\rightarrow \infty }{=}(-1)^{k}\frac{4\pi }{3^{k-2}}
\frac{k^{3}}{k!}E_{k}(-4)\,.  \label{nu-asymptotic-calc-4}
\end{equation}
Inserting the asymptotic expression for $E_{k}(-4)$ from 
eq. (\ref{E-k-4-asymptotic}) derived below, we obtain asymptotic formula
(\ref{nu-k-asymptotic}).

\subsection{Large-$k$ behavior of $E_{k}(-4)$}

Now we want to derive an asymptotic formula for $E_{k}(-4)$ at large $k$. At even
negative $N$ the problem of the anharmonic oscillator
(\ref{H-quartic-oscillator}) simplifies drastically. As was shown in 
Ref. \cite{DH-88} (see also \cite{PP-08}), energy $E(g,-4)$ is described by the cubic
equation
\begin{equation}
\mathcal{P}\left[ E(g,-4)\right] =0,\quad\mathcal{P}(E)\equiv E^{3}-4E-16g\,.
\label{E-cubic-eq}
\end{equation}
Series (\ref{E-expansion}) for $E(g,-4)$ has a finite radius of convergence
which is determined by the singularity $g_{0}$ of $E(g,-4)$ closest to the
point $g=0$. This singularity comes from degenerate roots of cubic equation
(\ref{E-cubic-eq}):
\begin{align}
\mathcal{P}(E_{0}) & =\mathcal{P}^{\prime}(E_{0})=0\,, \\
E_{0} & =-2/\sqrt{3}\,,\quad g_{0}=3^{-3/2}\,\,.
\end{align}
In the vicinity of this singular point
\begin{equation}
E(g,-4)\overset{g\rightarrow g_{0}}{=}E_{0}-\sqrt{8\cdot3^{-1/2}\left(
g_{0}-g\right) }\,.
\end{equation}
The small-$g$ expansion of this expression determines the large-$k$ behavior

\begin{equation}
E_{k}(-4)\overset{k\rightarrow\infty}{\longrightarrow}\frac{1}{3}
\sqrt{\frac{2}{\pi}}3^{3k/2}k^{-3/2}\,.  \label{E-k-4-asymptotic}
\end{equation}


\begin{thebibliography}{99}
\bibitem{BW-69} C.M. Bender, T.T. Wu, Phys. Rev. 184, 1231 (1969).

\bibitem{BW-73} C.M. Bender, T.T. Wu, Phys. Rev. D7, 1620 (1973).

\bibitem{Lipatov-77b} L.N. Lipatov, Sov. Phys. JETP 45, 216 (1977).

\bibitem{BLZ-77} E. Brezin, J. C. Le Guillou and J. Zinn-Justin, Phys. Rev.
D15, 1544 (1977).

\bibitem{KP-2002} D.I. Kazakov and V.S. Popov, J. Exp. Theor. Phys. 95, 581
(2002).

\bibitem{Suslov-05} I.M. Suslov, J. Exp. Theor. Phys. 100, 1188 (2005).

\bibitem{CMRSJ-07} E. Caliceti, M. Meyer-Hermann, P. Ribeca, A. Surzhykov
and U.D. Jentschura, Phys. Rept. 446, 1 (2007).

\bibitem{KNSCL-91} H. Kleinert, J. Neu, V. Schulte-Frohlinde, K.G.~Chetyrkin
and S.A. Larin, Phys. Lett. B272, 39  (1991), Erratum B319, 545  (1993)
[hep-th/9503230].

\bibitem{Kubyshin-84} Yu.A. Kubyshin, Theor. Math. Phys. 57, 1196 (1983).

\bibitem{MW-78} A.J. McKane and D.J. Wallace, J. Phys. A11, 2285 (1978).

\bibitem{MWA-1984} A.J. McKane, D.J. Wallace and O.F. de Alcantara Bonfim,
J. Phys. A17, 1861 (1984).

\bibitem{DKT-78} F.M. Dittes, Yu.A. Kubyshin and O.V. Tarasov, 
Theor. Math. Phys. 37,  879 (1978).

\bibitem{SZ-79} R. Seznec and J. Zinn-Justin, J. Math. Phys. 20, 1398 (1979).

\bibitem{Zinn-Justin-81} J. Zinn-Justin, J. Math. Phys. 22, 511 (1981).

\bibitem{ZJ-04} J. Zinn-Justin and U.D. Jentschura, Ann. Phys. (N.Y.) 313 (2004)
197, 269.

\bibitem{PP-08} P.V. Pobylitsa, \emph{Anharmonic oscillator, negative
dimensions and inverse factorial convergence of large orders to the
asymptotic form}, arXiv: 0807.5032 [quant-ph].

\bibitem{DH-88} G.V. Dunne and I.G. Halliday, Nucl. Phys. B308, 589 (1988).
\end{thebibliography}
\end{document}